\documentclass[conference]{IEEEtran}

\usepackage{amsmath}
\usepackage{amsthm}
\usepackage{amssymb}
\usepackage{amsfonts}
\usepackage{epsfig}
\usepackage{cmap}
\usepackage{color}

\newtheorem{theorem}{Theorem}
\newtheorem*{conjecture}{Conjecture}
\theoremstyle{remark}

\ifCLASSINFOpdf
 \else
 \fi

\begin{document}

\title{Spatial random multiple access with multiple departure}

\author{\IEEEauthorblockN{Sergey Foss}
\IEEEauthorblockA{Heriot-Watt University \\
EH14 4AS, Edinburgh, United Kingdom \\
and Sobolev Institute of Mathematics \\
and Novosibirsk State University \\
Email: s.foss@hw.ac.uk.}
\and
\IEEEauthorblockN{Andrey Turlikov, Maxim Grankin}
\IEEEauthorblockA{St.-Petersburg
University of \\ Aerospace Instrumentation \\ 67, B. Morskaya st.
St.-Petersburg \\
Email: turlikov@vu.spb.ru, m.a.grankin@gmail.com}}

\maketitle

\begin{abstract}
We introduce a new model of spatial random multiple access systems with
a non-standard departure policy: all arriving messages are distributed uniformly on
a finite sphere in the space, and when a successful transmission of a single message occurs, the transmitted message leaves the system together  with all its neighbours within a ball of a given radius centred at the message's location. We consider three classes of protocols: centralised protocols and decentralised protocols with either ternary or binary feedback; and analyse their stability. Further, we discuss some
asymptotic properties of stable protocols.
\end{abstract}

\IEEEpeerreviewmaketitle

\section{Introduction}
\label{sec:introduction}

In many real sensor systems, if one of sensors records an event, it is
also recorded by its close neighbours (for example, this happens with fire detectors). The main feature of such systems is: neighbouring sensors collect similar data, and given that a sensor has successfully transmitted the data to the information centre, its neighbours do not need to send
this information again (for example see Sensor Protocols for Information via Negotiation \cite{kulik2002negotiation}).

We introduce and analyse a new model that reflects this feature.
The model may be considered as a natural analogue of a random
multiple access system with a single transmission channel, infinitely many users and Poisson
input that was introduced in \cite{Tsy, Cap}. In this system, time is slotted and, within each time slot, users request transmission from a single transmission channel with some probabilities. Transmission is successful if there is only
one request, otherwise there is either collision (two or more requests) or an empty
slot (no requests). The ternary feedback (empty slot or success or collision)
is used to determine request probabilities in the next time slot. The authors of
\cite{Tsy, Cap} have designed various transmission protocols that may lead to stable
performance of the system.
In the particular case where at each given time instant all users take the decision to request transmission with {\it equal} probabilities, papers \cite{Haj, Mih} proposed  protocols that  are stable for any input rate below $e^{-1}$. These algorithms
perform efficiently also in the cases of binary feedbacks where
one cannot distinguish either an empty slot and successful transmission, or successful transmission and collision (see \cite{Meh,Par}), so the throughput capacity is again
$e^{-1}$.
The third case of binary feedback where empty slot and collision are indistinguishable is very different in nature and requires different ideas. It was studied in \cite{FHT}, where a new class
of so-called ``doubly randomised'' protocols was introduced and analysed; its stability was established in a particular case and several hypotheses were made. These hypotheses were verified in \cite{Che}: the throughput capacity of the new protocols is also $e^{-1}$.

\section{The Model and the Classes of Protocols}
\label{sec:Model_algorithm}

We consider a spatial variant of  a multi-access system introduced in
\cite{Tsy}.
There is an infinite number of users and a single transmission channel available
to all of them. Users exchange their messages using the channel. Time is slotted and all message
lengths are assumed to be equal to the slot length (and equal to one).

The input process of messages $\{ \xi_n\}$ is assumed to be i.i.d., having a
general distribution $F$. Here $\xi_n$ is the total number of messages arriving within
time slot $[n,n+1)$ (we call it ``time slot $n$'', for short).

All messages/users are located on a sphere of area 1 which is a surface of a ball
(of radius $R=1/\sqrt{4\pi}$). Each arriving message chooses its location
uniformly at random (and independently of everything else).

The system operates according to an ``adaptive ALOHA protocol'' that
may
be described as follows.
There is no coordination between the users, and at the beginning of time slot $n$ each message
present in the system is sent to the channel for transmission with probability $p_n$, independently of everything
else. So given that the total number of messages is $N_n$, the number of those sent to the channel,
$B_n$, has conditionally the Binomial distribution $B(N_n,p_n)$ (here $B_n \equiv 0$ if $N_n=0$). Let
$J_n=1$ if $B_n=1$ and $J_n=0$, otherwise. If $J_n=1$, then there is a successful transmission
within time slot $n$. Otherwise there is either an empty slot ($B_n=0$) or a collision of messages
($B_n \ge 2$), so there is no transmission.

In the classical setup of \cite{Tsy, Cap}, given successful transmission ($J_n=1$), the transmitted message leaves the system and all other messages stay in the system.
The novelty of the model under consideration is in the following. There is given
a number $r\ge 0$. Given a success, $J_n=1$, not only the successful message leaves the system but also all its ``neighbours'' that have Euclidian distance at most $r$ from it. Let $X_n$ be the set of locations of messages on the sphere
at time $n$ and $N_n=|X_n|$, its cardinality. Let $Y_n$ be the set of locations
of messages arriving within time slot $n$; clearly $\xi_n = |Y_n|$. Further, given
$J_n=1$, let
$W_n$ be the subset of $X_n$ of messages that are located within distance $r$ of the successful user and let $W_n=\emptyset$ if $J_n=0$. Then let $V_n=|W_n|$.

The
following recursion holds:
\begin{equation}\label{key1}
X_{n+1}=X_n-W_n+Y_n,
\end{equation}
where the operations $+$ and $-$ are viewed as set addition and set subtraction.
Then we get the recursion
\begin{equation}\label{key2}
N_{n+1}= N_n-V_n+\xi_{n}.
\end{equation}

Note that the model boils down to the classical one if $r=0$, however the cases
$r=0$ and $r>0$ are very different. On the other hand, if $r\ge R$, then we obtain
a simple model that accumulates messages and ``regenerates'' from time to time
by removing simultaneously all messages present.

A transmission {\it protocol} determines recursively transmission probabilities $p_1,p_2,\ldots$ based on an observable/known information. We consider
three types of models, (1) centralised models, (2) decentralised models with ternary feedback, and (3)
decentralised models with binary feedback ``success-nonsuccess''.
In all three cases, we consider Markov-type protocols: $p_n$ is a (random) number that depends on the history of the system only through its values that are available at time $n-1$.

In model (1), it is assumed that, at the beginning of any time slot $n$,
the past numbers $N_k$, $k=1,\ldots,n$ are known and may be used to determine
probability $p_{n}$. Here we consider the class ${\cal A}_1$ of protocols with the following transmission probabilities:
$p_n=c/N_n$ where the constant $c>0$ is the protocol parameter.

In model (2), the numbers $N_n$, $n=1,2,\ldots$ are not observable, and only values of past $\min (B_k,2), k<n$ are known. Following \cite{Haj}, we consider the class ${\cal A}_2$ of protocols
with the following dynamics: $p_{n+1}=c_1p_n$ if $B_n\ge 2$, $p_{n+1}=p_n$ if $B_n=1$ and $p_{n+1}=\min (1,c_2p_n)$ if $B_n=0$ where the constants $0<c_1<1<c_2$
are the protocol parameters.

In model (3), the numbers $N_n$, $n=1,2,\ldots$ are not observable, and only values of past $J_k, k<n$ are known (so one cannot distinguish $B_n=0$ and $B_n\ge 2$).
Here the class of protocols ${\cal A}_3$ has a more complex form and is
determined by the constant $C$, positive functions $h(x)\uparrow \infty$ and $\varepsilon (x)\downarrow 0$ as $x\to\infty$, independent i.i.d. sequence $\{I_n\}$ with ${\mathbf P} (I_n=1)={\mathbf P} (I_n=0)=1/2$ and the auxiliary sequence $\{K_n\}$  as follows.
Given $K_n$, we let
\begin{equation*}
p_n=
\begin{cases}
(1-\varepsilon_h(K_n)) /K_n & \text{if}\ \  I_n=0,\\
1/K_n & \text{if} \ \ I_n=1,
\end{cases}
\end{equation*}
and then define $K_{n+1}$ by
\begin{equation*}
K_{n+1}=
\begin{cases}
K_n + C & \text{if}\ \  J_n=0,\\
K_n + h(K_n) & \text{if} \ \ J_n=1 \ \ \text{and} \ \ I_n=0,\\
\max (K_n - h(K_n),1) & \text{if} \ \ J_n=1 \ \ \text{and} \ \
I_n=1.
\end{cases}
\end{equation*}

\section{Stability}
\label{sec:Centr}

In this Section, we assume $r>0$.
We say that a transmission protocol is {\it stable} if the underlying Markov
chain is {\it Harris ergodic}: there exists a unique stationary distribution and,
for any initial fixed value and as time frows, the distribution of the Markov chain converges in the total variation norm to the stationary one.

\begin{theorem}\label{Th-1} Assume that  $a :={\mathbf P} (\xi_1=0)>0$. Then
any protocol from the class ${\cal A}_1$ is stable.
\end{theorem}

{\sc Proof.} Split the sphere into a finite number, say $M$, of non-overlapping sets $A_j$, $j=1,\ldots,M$, each of which has the diameter at most $r$ (the diameter of a set is the maximal distance between its points).

Fix $c>0$ and let
$b = \min_k c(1-c/k)^{k-1}>0$ be the minimal probability of successful transmission over all $k$. For any $n$, introduce the event
$$
D_n = \bigcap_{i=0}^{M-1} \left(\{|X_{n+i}|=0\}\cup \{|X_{n+i}|>0,B_{n+i}=1\}\right)
\cap \{\xi_{n+i}=0\}.
$$

Notice that the probability of the event $D_n$ is a least $q:= a^Mb^M >0$.
Indeed, the probability of having no arrivals within $M$ consecutive time slots is $a^M$ and, for any $i$,
\begin{eqnarray*}
&&{\mathbf P}\left(\{|X_{n+i}|=0\}\cup \{|X_{n+i}|>0,B_{n+i}=1\} \ | \ E \right) \\
&\ge &
{\mathbf P} (|X_{n+i}|=0 \ | \ E) +  b{\mathbf P}(|X_{n+i}|>0 \ | \  E) \\
&\ge & b,
\end{eqnarray*}
for any event $E$ determined by the history up to time $n+i-1$.

Next, given the event $D_n$ occurs, we have
$X_{n+M}=\emptyset$. Indeed, since there are no arrivals, if $X_{n+i-1}$ is the empty set, $X_{n+i}$ is empty too.
It is left to show that, given the event
$$
\widehat{D}_n :=
\bigcap_{i=0}^{M-1} \{|X_{n+i}|>0,B_{n+i}=1\}
$$
occurs, we get $X_{n+M}=\emptyset$. To see this, we recall that a successful transmission of a message in a given time slot
imples not only its removal, but also removal of all its radius-$r$ neighbours. In particular, this means that if the transmitted message
was located in, say, set $A_j$, then all messages from this set are removed together with it. Therefore, given the event $\widehat{D}_n$ occurs,
at least one (and, in fact, exactly one) set from $\{A_j\}_{j=1}^M$ is cleared of messages at each time $t=n,\ldots,n+M-1$.
Thus, given the event $D_n$, all sets $A_j$ are cleared of messages by time
$n+M$ (and, therefore, the whole system is cleared). Then the event $D_n$ is regenerative for the Markov chain $\{X_n\}$. Let
$$
\gamma = \min \{ n \ : \ \mbox{event} \ D_{nM} \ \mbox{occurs}\}.
$$
Then, for any initial value $X_0$,
$$
{\mathbf P} (\gamma >n) \le {\mathbf P} \left(\cap_{i=1}^n \overline{D}_{iM}\right) \le (1-q)^n,
$$
here $\overline{D}_{iM}$ is teh complement of the event $D_{iM}$.
Then the Markov chain is {\it uniformly ergodic} and regenerates geometrically fast. It is clearly aperiodic and, therefore, its distribution converges
geometrically fast to the stationary distribution.

{\it Remark.} The result of Theorem \ref{Th-1} holds also without the assumption $a>0$, but then there is no regeneration at the empty set and the proof becomes much
more lengthy. One can show that the Markov chain is still Harris ergodic using the A.A.Borovkov's ``renovation theory'' (see, e.g., the overview paper \cite{FK}).

We believe that similar results should hold for the decentralised protocols, but at
the moment
are able to prove only a weaker statement.

\begin{theorem}\label{Th-C}
Assume that ${\mathbf E}\log \max (1,\xi_1)$ is finite. Then\\
(1) there exists a stable protocol in the class
${\cal A}_2$;\\
(2) there exists a stable protocol in the class ${\cal A}_3$.
\end{theorem}

Our proof of Theorem \ref{Th-C} is much more lenthy. We consider protocols that are described in the previous Section, with taking a particular choice of their parameters and parametric functions. The proof is based on a generalised version of the Foster criterion (see, e.g., \cite{FK}), using a certain  logarithmic test function.

\section{Mean Delay}

\label{sec:Delay}

Assume $\{\xi_n\}$ to be Poisson random variables with finite mean $\lambda >0$. Let $D_{r,\lambda}$ be
the sojourn time (delay) in the system of a typical message in the stationary regime
and ${\mathbf E} D_{\lambda}$ its mean. Let $S_r$ be the area of the circle of radius $r$ on the sphere.

Consider the centralised model (1) with transmission probabilities $p_n=1/N_n$.
We know that ${\mathbf P} (B_n=1 \ | \ N_n=k) \uparrow e^{-1}$ as $k\to\infty$.
We formulate the following conjecture.

\begin{conjecture} Fix $r\in (0,2R)$. There exists a positive finite $Z_r$ such that,
for all sufficiently large $\lambda$, we have
$$
{\mathbf E}D_{r,\lambda} \le Z_r 
$$
and further
$$
\lim_{\lambda\to\infty}
{\mathbf E}D_{r,\lambda}= Z_r 
$$
Then the next problem is to identify $Z_r$.
\end{conjecture}
Simple observations show that $Z_r$ possesses the following upper bound: $Z_r<e/S_r$, for any $r<2R$.

Our conjecture is supported by simulations; see Figure \ref{delay_vs_lamda}
below.

\begin{figure}[h]
	\centering
	\includegraphics[width=1\columnwidth]{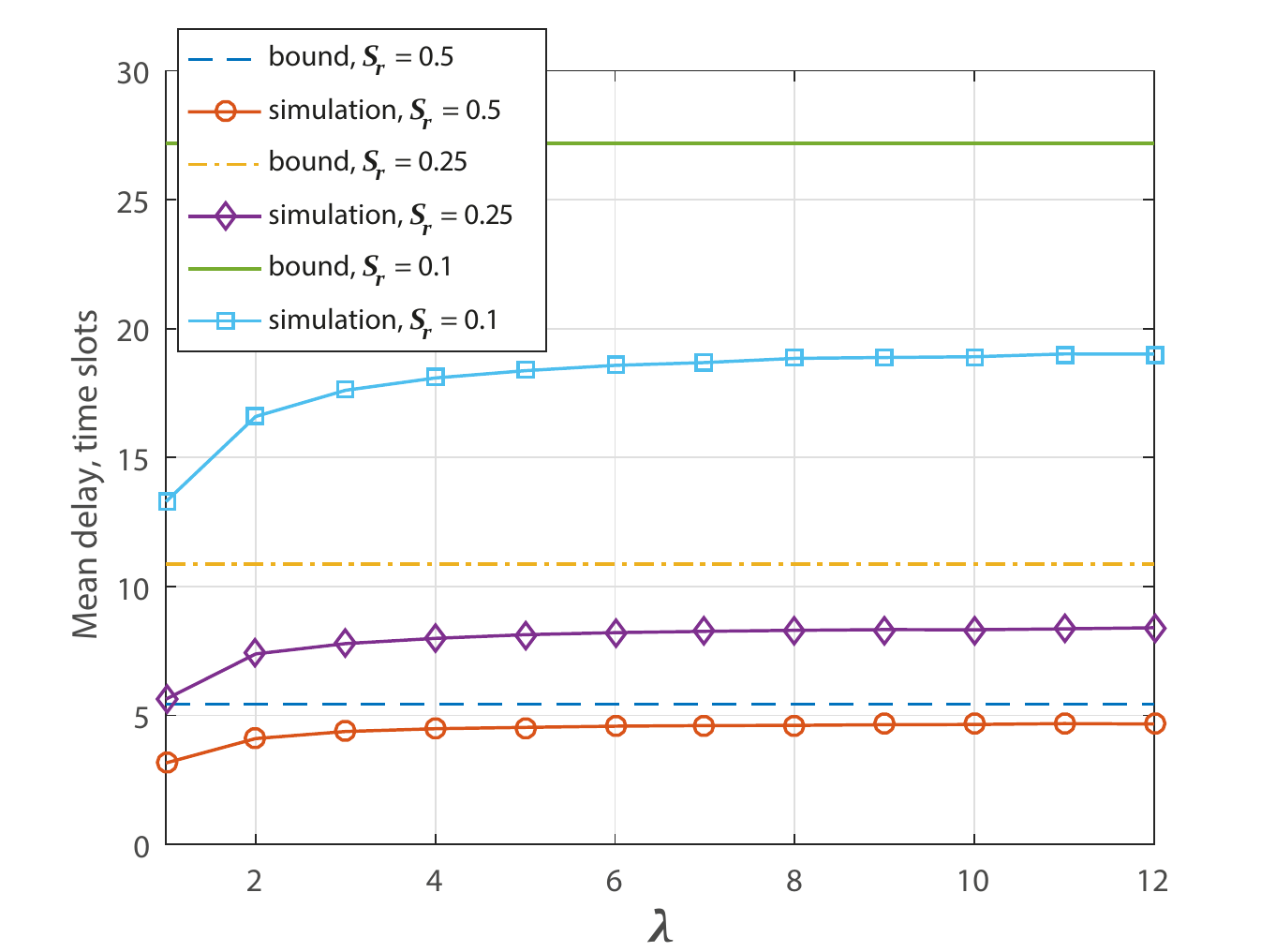}
	\caption{Mean delay vs $\lambda$}
	\label{delay_vs_lamda}
\end{figure}


\begin{figure}[h]
	\centering
	\includegraphics[width=1\columnwidth]{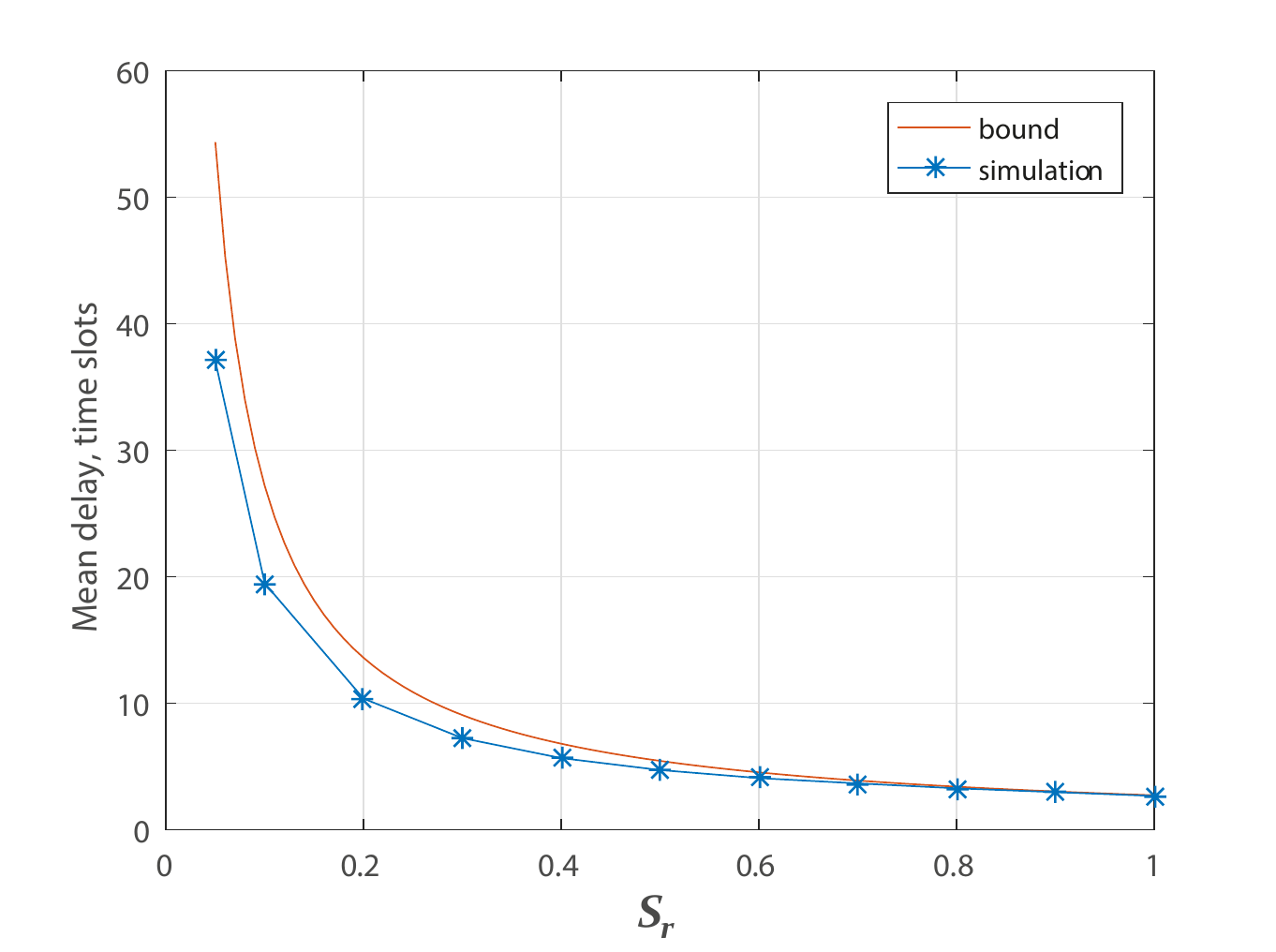}
	\caption{Mean delay vs $S_r$}
	\label{delay_vs_Sr}
\end{figure}

Figure \ref{delay_vs_Sr}
shows that  an increase in $r$ implies a decrease in the mean delay and leads to an increase in the bound accuracy. 


\section{Conclusions}

We have introduced a natural analogue of the classical random multiple access
model with a single transmission channel and considered several classes of randomised adaptive protocols where all users send a transmission request with equal probabilities.

We have shown that our system is stable under mild assumptions (see Theorems 1 and 2).  Notice that the classical model is stable only if the input rate is finite and is smaller than $e^{-1}$.

Further, we have formulated the conjecture that, in the centralised model with a Poisson input with a sufficiently large intensity, the mean delay is bounded from above by a finite number. A further problem is to find this number.

\section{Acknowledgment}

The authors (Turlikov and Grankin) are supported by scientific project  $N\textsuperscript{\underline{o}}$ 8.8540.2017 ``Development of data transmission algorithms in IoT systems with limitations on the devices complexity'' and (Foss) by Grant 1030/GF4 of Ministry of Education and Science of Kazakhstan.


\end{document}